# Thermal Transport at the Nanoscale: A Fourier's Law vs. Phonon Boltzmann Equation Study

J. Kaiser,[1,a)] T. Feng,[2] J. Maassen,[3] X. Wang,[4] X. Ruan,[2] and M. Lundstrom[4]

[1]*Department of Electrical Engineering and Information Science, Ruhr-University Bochum, D-44780 Bochum, Germany*

[2]*School of Mechanical Engineering, Purdue University, West Lafayette, Indiana 47907, USA*

[3]*Department of Physics and Atmospheric Science, Dalhousie University, Halifax, Nova Scotia B3H 4R2, Canada*

[4]*School of Electrical and Computer Engineering, Purdue University, West Lafayette, Indiana 47907, USA*

Steady-state thermal transport in nanostructures with dimensions comparable to the phonon mean-free-path is examined. Both the case of contacts at different temperatures with no internal heat generation and contacts at the same temperature with internal heat generation are considered. Fourier's Law results are compared to finite volume method solutions of the phonon Boltzmann equation in the gray approximation. When the boundary conditions are properly specified, results obtained using Fourier's Law *without modifying the bulk thermal conductivity* are in essentially exact quantitative agreement with the phonon Boltzmann equation in the ballistic and diffusive limits. The errors between these two limits are examined in this paper. For the four cases examined, the error in the apparent thermal conductivity as deduced from a correct application of Fourier's Law is less than 6%. We also find that the Fourier's Law results presented here are nearly identical to those obtained from a widely used ballistic-diffusive approach, but analytically much simpler. Although limited to steady-state conditions with spatial variations in one dimension and to a gray model of phonon transport, the results show that Fourier's Law can be used for linear transport from the diffusive to the ballistic limit. The results also contribute to an understanding of how heat transport at the nanoscale can be understood in terms of the conceptual framework that has been established for electron transport at the nanoscale.

## I. INTRODUCTION

The treatment of heat transport in nanostructures with dimensions comparable to the phonon mean-free-path is a problem of both fundamental and practical interest.[1-3] Beginning with the work of Joshi and Majumdar[4], much has been learned about thermal transport at the nanoscale (as reviewed, for example, in Chapter 7 of Ref. 3). Rigorous techniques, such as molecular dynamics simulations[5] or solving the phonon Boltzmann Transport Equation (BTE) directly[6], have been essential in understanding nanoscale heat transport, but physically sound, analytically compact, and computationally efficient approaches are also much-needed. Majumdar showed how to use Fourier's Law at the nanoscale by replacing the thermal conductivity with a size-dependent, apparent thermal conductivity.[7] Chen and Zeng showed that the direct use of Fourier's Law without modifying the thermal conductivity can produce quite accurate results, at least for one-dimensional problems.[8] The key is to use appropriate (temperature-jump) boundary conditions. Because of the need for computationally efficient approaches, extensions of Fourier's Law have been considered by many researchers (e.g. see Refs. 9-11 and references therein).

---

1. Electronic mail: jan.kaiser@rub.de

In this paper, we examine the use of the unmodified Fourier's Law at the nanoscale, but with special boundary conditions at the contacts. In this regard, the recent work of Peraud and Hadjiconstantinou[10] is relevant. Peraud and Hadjiconstantinou present asymptotic expansion solutions of the Boltzmann equation focusing on small Knudsen numbers.[10] Our paper examines the use of Fourier's Law across the entire diffusive to ballistic spectrum. Peraud and Hadjiconstantinou show that the zeroth order solution is the classic Fourier Law solution with fixed temperatures at the boundaries, but the first and second order solutions involve temperature jumps at the boundaries. Their analysis shows that at least up to second order, the thermal conductivity in the bulk is the unmodified bulk conductivity— even in small structures. They point out that there is no justification for introducing an effective thermal conductivity in small structures; the reduction of thermal transport is due to the temperature jump boundary conditions, not to a reduced thermal conductivity. These are the same conclusions that we arrive at. The difference is that Peraud and Hadjiconstantinou treat the full BTE by asymptotic expansion and focus on the small Knudsen number regime. In contrast, we first simplify the BTE (the McKelvey-Shockley equations) and then show that these equations lead without further approximation to Fourier's Law and that temperature jump boundary conditions arise naturally from using physically correct boundary conditions for the BTE itself. Peraud and Hadjiconstantinou introduce kinetic boundary layer functions to treat the non-linear temperature profiles near the boundaries. We ignore these boundary layers and treat the entire region inside the contacts with Fourier's Law. For moderate Knudsen numbers, our solution is less accurate, but in the diffusive limit and the ballistic limit (which is not examined in Ref. 10), our solution is exact. The main conclusion of our work agrees with that of Peraud and Hadjiconstantinou – that one should use the unmodified Fourier's Law inside a nanostructure, but the boundary conditions must be modified to a jump type boundary condition.

This paper builds on the work of Maassen and Lundstrom[12] who extended the work by Chen and Zang[8] by introducing a consistent definition of temperature at the nanoscale (analogous to the way that electrochemical potentials are defined at the nanoscale[15]) and by showing how to derive Fourier's law without assuming local thermodynamic equilibrium. The work reported here extends that in Ref. 12 by considering the important case of nanostructures with internal heat generation and by carefully comparing results obtained from Fourier's Law to numerical solutions to the phonon BTE assuming a simple, steady-state, gray model. This comparison confirms that Fourier's Law produces exact solutions in the diffusive *and* ballistic limits, and it quantifies the errors between these limits. The Fourier's Law analysis presented here also provides new insights into heat transport in nanostructures with internal heat generation, such as how to describe temperature in terms of the temperatures of forward and reverse fluxes and the fact that even under diffusive conditions, temperature jumps can occur at contacts. We show that the critical issue is not the validity of Fourier's Law itself, but rather the boundary conditions to apply to the heat equation.

The six model structures shown in Fig. 1 were recently examined by Hua and Cao[16] who used a simple gray model and solved the steady-state phonon BTE by Monte Carlo techniques. Structures (a) and (b) in Fig. 1 are infinite in the y- and z-directions, so transport is one-dimensional. Structures (c) and (d) are thin in the y-direction and assume diffusive scattering at the boundaries. Structures (e) and (f) are nanowires with diffusive boundary scattering. In this paper, we consider structures (a) – (d) using material parameters appropriate to silicon at room temperature (thermal conductivity, $\kappa_{bulk} = 160\ W/$(mK), specific heat, $C_V = 1.63 \times 10^6$ J/(m³ K), sound velocity, $v_s = 6400$ m/s, $\tau = 7.19$ ps, which results in a phonon mean-free-path of $\Lambda = 46.0$ nm). Structures (e) and (f) of Fig. 1 are discussed in the Supplementary Information. We will compare results obtained from Fourier's Law to those obtained from a finite volume method solution to the phonon BTE.[17] In the Supplementary Information, we compare our solution to the results of Hua and Cao obtained by solving the same gray model phonon BTE using Monte Carlo techniques.[16,18]

The paper is organized as follows. In Sec. II, the use of Fourier's Law at the nanoscale[12-14] is briefly reviewed. Results are presented in Sec. III, and the results are discussed in Sec. IV, which also discusses the source of the differences in the two methods observed in the quasi-ballistic regime. Section V summarizes the conclusions of the paper.

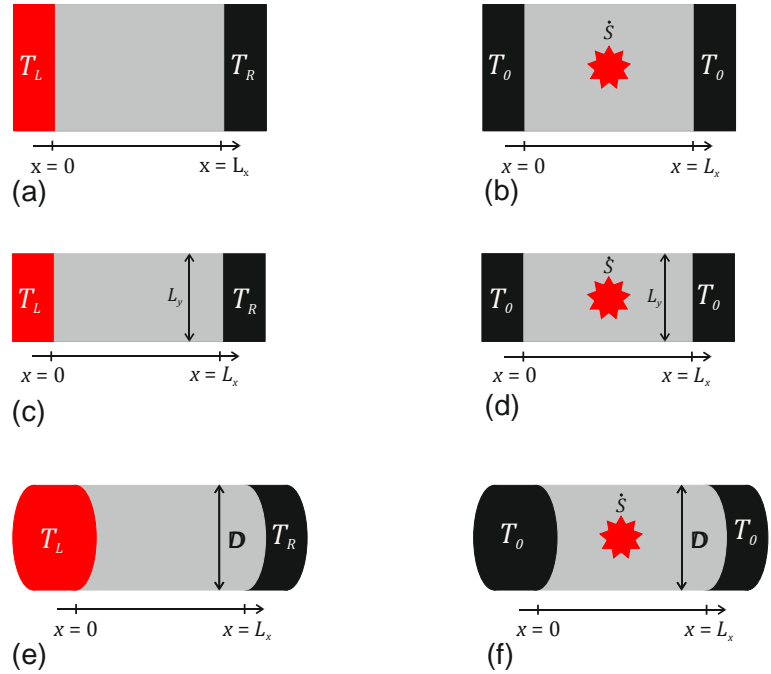

FIG. 1. Model structures examined: with no internal heat source and contacts at different temperatures (a, c, e) and with internal heat source and contacts at the same temperature (b, d, f). (After Hua and Cao[16])

## II. FOURIER'S LAW AT THE NANOSCALE

The use of Fourier's Law at the nanoscale has been discussed in Ref. 12-14; only a brief summary for the steady-state condition of interest in this paper is provided here. More details are provided in the Supplementary Information and in Ref. 12 (see also the Supplementary Information for Ref. 12).

We begin with the steady-state flux equations as written by Shockley:[19, 20]

$$\frac{dF_Q^+(x)}{dx} = -\frac{F_Q^+(x)}{\lambda} + \frac{F_Q^-(x)}{\lambda} + \frac{\dot{S}}{2} \tag{1a}$$

$$\frac{dF_Q^-(x)}{dx} = -\frac{F_Q^+(x)}{\lambda} + \frac{F_Q^-(x)}{\lambda} - \frac{\dot{S}}{2}, \tag{1b}$$

where $F_Q^+(x)$ is the forward-directed heat flux, $F_Q^-(x)$ the negative-directed heat flux, $\lambda$ the "mean-free-path for backscattering" (see the appendix in Ref. 20 and Ref. 21). The term, $\dot{S}$, is a heat generation term assumed to be spatially uniform in this paper. For 3D isotropic phonons, the mean-free-path for backscattering is related to the conventional mean-free-path, $\Lambda = v_s \tau$,[20, 21]

$$\lambda = \frac{4}{3}\Lambda. \tag{2}$$

Temperatures can be associated with the forward and reverse fluxes[12]

$$F_Q^+ = v_x^+ \frac{C_V}{2} T^+ \tag{3a}$$

$$F_Q^- = v_x^+ \frac{C_V}{2} T^-, \tag{3b}$$

where $v_x^+ = v_s/2$ is the average +x-directed velocity, $C_V$ is the specific heat per unit volume, and $v_s$ is the sound velocity. $T^+$ and $T^-$ should be understood to be temperatures relative to a background temperature, $T_0$.[12] Small deviations in temperature are assumed so that the specific heat can be treated as a constant. Our use of two different temperatures for the forward and reverse streams has been discussed in Ref. 12 and is analogous to how the electrochemical potential has been defined at the nanoscale.[15] Finally, we note that the flux equations can be derived from the Boltzmann Transport Equation. They can be regarded as a type of differential approximation to the Equation of Phonon Radiative Transport (ERPT) in which we integrate separately over the forward and reverse directions rather than over all directions.[4, 7, 22] In the Supplementary Information, we relate the flux equations to the ERPT.

By adding and subtracting eqns. (1a) and (1b), we find

$$\frac{dF_Q}{dx} = \dot{S} \tag{4a}$$

$$F_Q = -\kappa_{bulk} \frac{dT}{dx}, \tag{4b}$$

where

$$F_Q(x) = F_Q^+(x) - F_Q^-(x) \tag{5}$$

is the net heat flux,

$$\kappa_{bulk} = \frac{v_x^+ \lambda}{2} C_V = \frac{1}{3} v_s \Lambda C_V \tag{6}$$

is the thermal conductivity, and

$$T = (T^+ + T^-)/2 \tag{7}$$

is the average temperature of the forward and reverse heat fluxes. Equations (4a) and (4b) lead to a steady-state heat diffusion equation,

$$\frac{d^2 T}{dx^2} = -\frac{\dot{S}}{\kappa_{bulk}}, \tag{8}$$

that is mathematically identical to eqns. (1). Equations (1) apply from the ballistic to diffusive limits. Accordingly, eqn. (8) also applies from the ballistic to diffusive limits. The thermal conductivity, $\kappa_{bulk}$, is not size dependent (unless we bring in surface roughness scattering as discussed later for thin films). The fact that Fourier's Law and the heat diffusion equation can be used from the diffusive to ballistic limits with the bulk thermal conductivity has been discussed in Ref. 12. We must, however, be careful about the boundary conditions when using eqn. (8).[12] We shall see that a size dependent "apparent thermal conductivity" results when the proper boundary conditions are used (see eqn. (15) below). Peraud and Hadjiconstantinou reached the same conclusion.[10]

The boundary conditions for the phonon BTE are the incident heat fluxes from the two contacts. (Ideal black body contacts are assumed.) The temperatures at the two ends of the film are a result of the calculation and can only be imposed in the diffusive limit. As shown in Ref. 12, when the correct boundary conditions are used, temperature jumps can occur – even for ideal contacts. The temperatures at the two contacts can be written as

$$T(0^+) = T_L - \Delta T(0) \tag{9a}$$

$$T(L_x^-) = T_R + \Delta T(L_x), \tag{9b}$$

where $T_L$ is the temperature of the left contact and $T_R$ is the temperature of the right contact. The temperature jumps can be shown to be the product of the net heat flux and one-half of the ballistic thermal resistance[12]

$$\Delta T(0) = F_Q(0) \frac{R_B A}{2} \tag{10a}$$

$$\Delta T(L) = F_Q(L) \frac{R_B A}{2}, \tag{10b}$$

where $A$ is the cross-sectional area and

$$R_B A = \frac{2}{C_V v_x^+} \quad (11)$$

is the ballistic thermal resistance. Note that $R_B$ is a fundamental thermal boundary resistance for the assumed ideal, reflectionless (black) contacts. Real contacts would have additional interface resistance.

To summarize, we solve eqn. (8) with boundary conditions specified by eqns. (9) – (11). After solving for $T(x)$, the directed temperatures can be obtained from

$$T^+(x) = T(x) + F_Q(x) R_B A/2 \quad (12a)$$

$$T^-(x) = T(x) - F_Q(x) R_B A/2 . \quad (12b)$$

Use of these equations will be illustrated as we discuss the model structures shown in Fig. 1.

Finally, we note that the specification of boundary conditions in terms of the ballistic resistances simplifies the calculations and may be useful in other contexts as well. For example, it is well-known that thermal transport can be simulated using an electrical network analogy.[23] Using the equivalent circuit in Fig. 2 below, all of the steady-state, transient, and small-signal results presented in Ref. 12-14 (as well as all of the results to be reported in this paper) can be obtained by circuit simulation. This equivalent circuit describes thermal transport from the ballistic to diffusive limits and is identical to the standard equivalent circuit for thermal transport except for the addition of one-half of the ballistic resistance at each of the two contacts.[23]

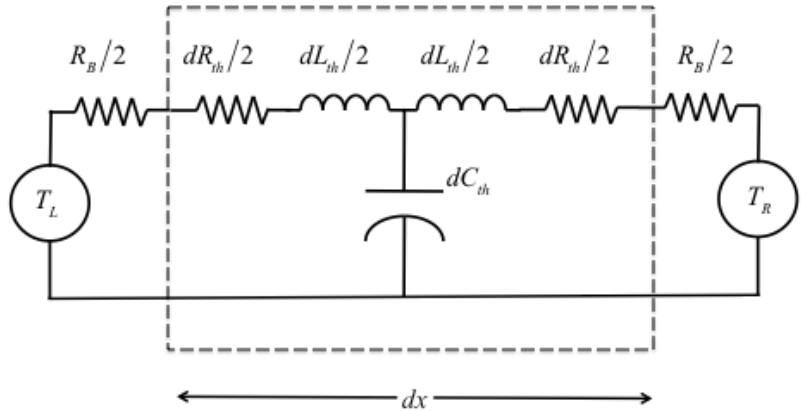

FIG. 2. Equivalent circuit for the treatment of thermal transport from the ballistic to diffusive limits. The circuit simply adds ballistic contact resistances to the standard, diffusive equivalent circuit [23]. Here $R_{th} = dx/(\kappa_{bulk} A)$, $dL_{th} = \tau_Q\, dR_{th}$, where $\tau_Q$ is a scattering time in the Catteneo equation [13], and $dC_{th} = A\, C_V\, dx$. For a typical problem, the structure would be divided into several sections to spatially resolve the temperature profiles, but the ballistic resistors should only be included at the two contacts (i.e. there would be several sections like that in the dashed rectangle, but only two ballistic resistors).

## III. RESULTS

In this section, four of the cases illustrated in Fig. 1 are considered. In each case, we present the Fourier's Law solution and compare it to finite volume method (FVM) solutions of the BTE.[17]

### A. Cross-plane nanofilm with no internal heat generation

Consider first the case of Fig. 1a, where the contacts are at different temperatures, and there is no internal heat source. The length in the y-direction is assumed to be long enough so that lateral boundaries have no influence on the phonon transport. According to (8) with $\dot{S} = 0$, the temperature profile is linear, so we find

$$F_Q = \kappa_{bulk} \left( \frac{T_L - \Delta T - (T_R + \Delta T)}{L_x} \right). \tag{13}$$

Using (10) for $\Delta T = \Delta T(x = 0) = \Delta T(x = L_x)$, we find

$$F_Q = \kappa_{app} \left( \frac{T_L - T_R}{L_x} \right), \tag{14}$$

where

$$\kappa_{app} = \frac{\kappa_{bulk}}{1 + \lambda/L_x} = \frac{\kappa_{bulk}}{1 + 4Kn_x/3} \tag{15}$$

is the apparent thermal conductivity, which differs from the bulk thermal conductivity, $\kappa_{bulk}$, due to quasi-ballistic phonon transport in the x-direction. The Knudsen number, $Kn_x$, is defined as $Kn_x \equiv \Lambda/L_x$.

The temperature profile is

$$T(x) = (T_L - \Delta T)\left(1 - \frac{x}{L_x}\right) + (T_R + \Delta T)\left(\frac{x}{L_x}\right) \tag{16}$$

and the temperature jumps are obtained from (10) as

$$\Delta T = \frac{1}{2}\left(\frac{\lambda}{\lambda + L_x}\right)(T_L - T_R) = \frac{\mathcal{T}}{2}(T_L - T_R) = \frac{1}{2}\left(\frac{T_L - T_R}{1 + 3/(4Kn_x)}\right). \tag{17}$$

The temperature jump is one-half the phonon transmission, $\mathcal{T}$, times the difference in the contact temperatures. The last expression on the RHS is eqn. (27) in Ref. 18. The result has been obtained a number of times in the past using a variety of methods; it results here from a simple solution to the heat equation using Fourier's Law and appropriate boundary conditions. Note that eqn. (17) applies in both the ballistic to diffusive limits as well as in between these limits.

The normalized temperature profiles for several different Knudsen numbers are plotted in Fig. 3, which compares the Fourier's Law solution as given by eqn. (16) to FVM BTE simulations. In the diffusive limit, $T(x)$ varies linearly from $T_L$ to $T_R$ and both solutions agree. Near the ballistic limit ($Kn_x = 100$ in Fig. 3), $T(x) = (T_L + T_R)/2$, and Fourier's Law gives the correct answer. Figure 3 shows differences in the quasi-ballistic regime ($1 < Kn_x < 10$), which get smaller for $Kn_x \ll 1$ and for $Kn_x \gg 10$. We conclude that for case (a) in Fig. 1 (which is much like the case treated in Ref. 12), Fourier's Law with correct boundary conditions in the heat equation provides an exact description of ballistic and diffusive transport and an approximate solution between those limits.

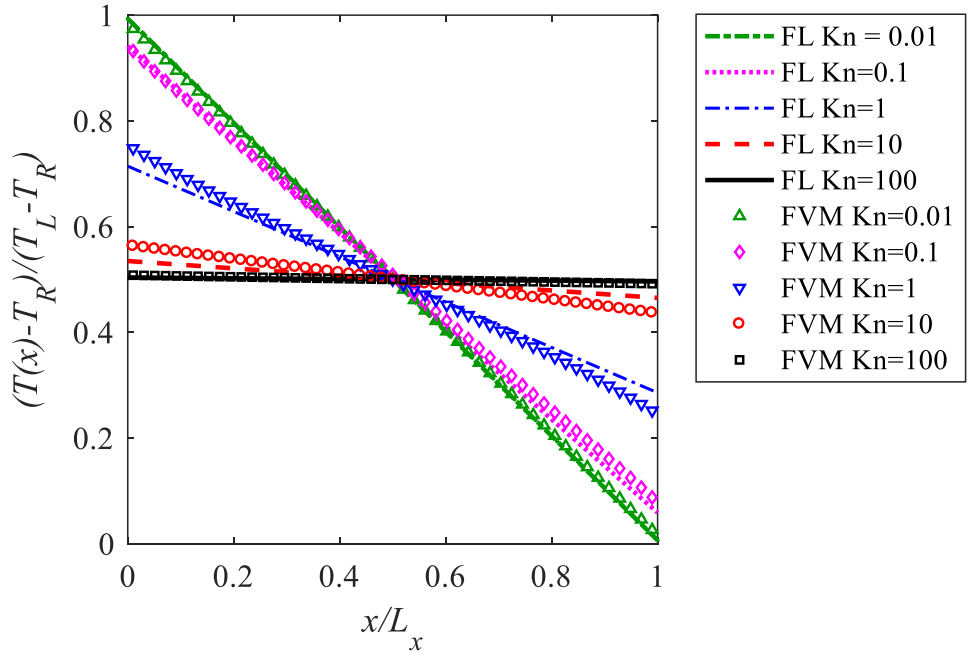

FIG. 3. Normalized temperature profile $(T(x) - T_R)/(T_L - T_R)$ vs. normalized distance, $x/L_x$, for cross-plane heat transport with no internal heat generation (Fig. 1a). Several different Knudsen numbers are shown. Lines are the result of Fourier's Law, and the symbols are FVM solutions of the phonon BTE.

Figure 4, a plot of the normalized temperature jump vs. Knudsen number, shows the differences between the Fourier's Law solution and the FVM BTE solution more clearly. The differences first increase as $Kn_x$ increases and then decrease as $Kn_x$ continues to increase towards the ballistic limit. The error vs. $Kn_x$ is also plotted in Fig. 4, which shows that the maximum error occurs at $Kn_x = 2.3$ and is less than 4%. Fourier's Law is exact in the ballistic and diffusive limits (small numerical errors are seen in the FVM solution because the BTE becomes stiff in the diffusive limit).

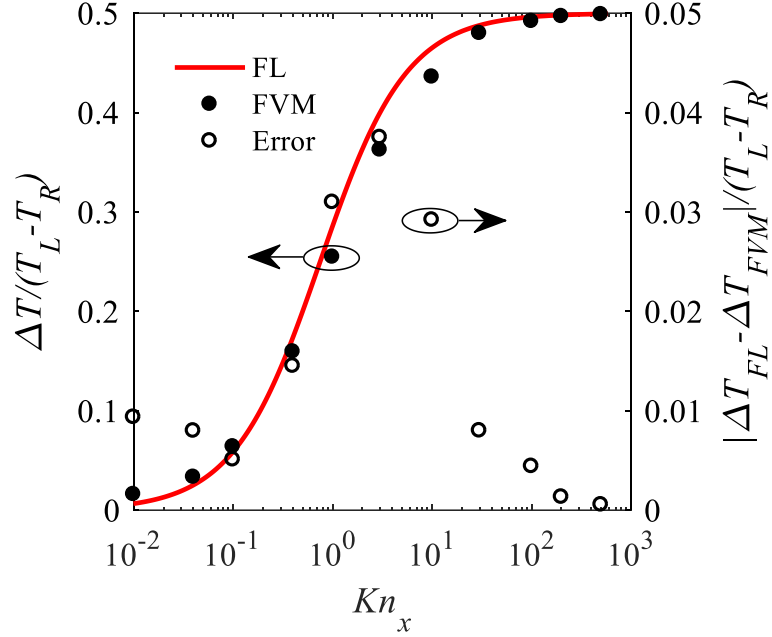

FIG. 4. The left axis shows the normalized temperature jump, $\Delta T(x=0)/(T_L - T_R)$ vs. $Kn_x$ for cross-plane thermal transport with no internal heat generation (Fig. 1a). The Fourier's Law solution (line) is from eqn. (10a), and filled symbols are the FVM solutions to the phonon BTE. The empty symbols belong to the right axis and show the error, $|\Delta T_{FL} - \Delta T_{FL}|/(T_L - T_R)$, between both solutions.

## B. Cross-plane nanofilm with internal heat generation

We turn next to the case shown in Fig. 1b, cross-plane heat transport with a uniform internal heat generation and both contacts at the same temperature, $T_L = T_R = T_0$. This problem has been considered by Zeng and Chen[24] and by Bulusu and Walker[25], who solved the one-dimensional phonon BTE exactly, and recently by Hua and Cao[16], who solved the two-dimensional phonon BTE by Monte Carlo simulation.

Equation (8) can be solved to find

$$T(x) = \left(\frac{\dot{S}}{2\kappa_{bulk}}\right)(L-x)x + T_b, \qquad (18)$$

where we are careful not to assume $T_b = T_0$. The temperatures at the boundaries are obtained from eqns. (10) with

$\Delta T(0) = -(T_b - T_0) = -\Delta T(L_x)$.

We find

$$|\Delta T| = T_b - T_0 = \left(\frac{\dot{S}L}{2}\right)\frac{1}{C_v v_x^+}. \qquad (19)$$

The maximum temperature occurs at $x = L_x/2$. From eqns. (18) and (19), we find

$$\frac{\delta T}{|\Delta T|} = \frac{T(L_x/2) - T_b}{T_b - T_0} = \frac{L_x}{2\lambda} = \frac{1}{8Kn_x/3}, \tag{20}$$

where $\delta T = T(x = L_x/2) - T_b$. The solution is sketched in Fig. 5. It is interesting to note that the temperature jumps at the boundaries do not depend on the mean-free-path, but the rise in temperature inside the film does. The more diffusive the sample, the higher the peak temperature. The more ballistic the sample, the lower the peak temperature until the ballistic limit is reached where $T(x) = T_b$. Note that a traditional Fourier's Law solution to this problem (i.e. assuming that $T(0) = T(L) = T_0$, would be incorrect even for when $L_x \gg \Lambda$, but the error would be small because the temperature jump at the boundary, $\Delta T$, would be much less than the temperature rise inside the structure, $\delta T$.

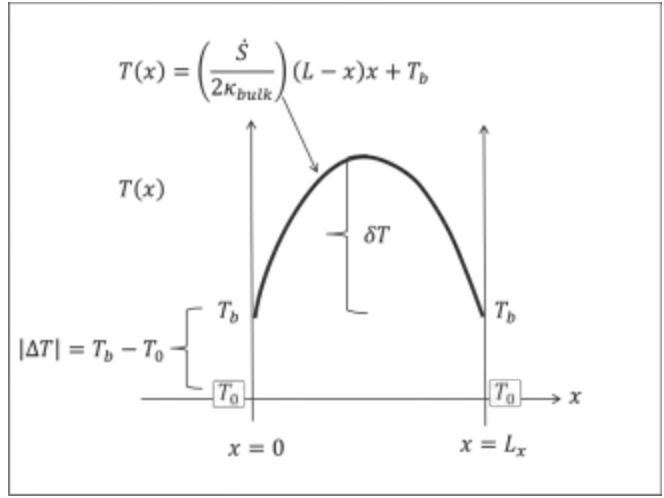

FIG. 5. Sketch of the solution, $T(x)$, for a sample with internal heat generation and two contacts at the same temperature.

Figure 6 plots the normalized temperature, $(T(x) - T_b)/(T_b - T_0)$ vs. normalized distance, $x/L_x$ for several different Knudsen numbers and compares our Fourier's Law solution to FVM BTE simulations.[17] As $Kn_x \to 0$, $(T(x = L_x/2) - T_b)/(T_b - T_0) \to \infty$, and the agreement in the diffusive limit is excellent. As $Kn_x \to \infty$, $(T(x = L_x/2) - T_b)/(T_b - T_0) \to 0$, and the agreement in the ballistic limit is excellent. As for the example with no internal heat generation, errors occur between the ballistic and diffusive limits. Finally, we note that although much simpler in form, the Fourier Law solution, eqns. (18) and (19), gives results that are essentially identical to the ballistic-diffusive solution presented as eqn. (23) in Hua and Cao.[16, 26]

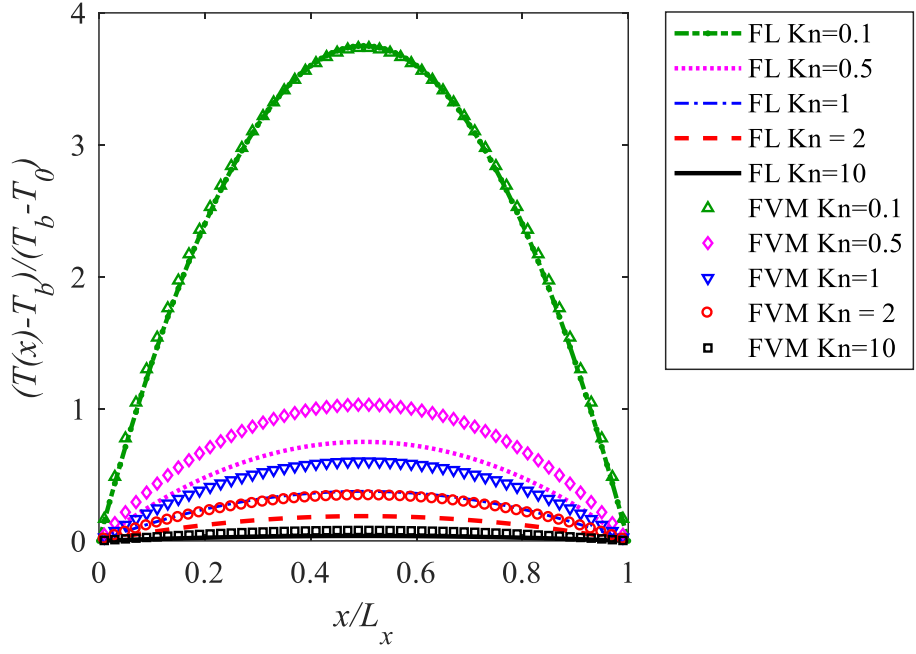

FIG. 6. Nanofilm (cross-plane) with internal heat generation (Fig. 1b). Plot of $(T(x) - T_b)/(T_b - T_0)$ vs. $x/L_x$ for several different values of $Kn_x$. Lines are Fourier's Law solutions and symbols are FVM solutions of the phonon BTE.

Figure 7, a plot of the normalized temperature rise, $\delta T/\Delta T$, in the center of the film as given by eqn. (20) vs. $Kn_x$ shows the differences between our Fourier Law solutions and the FVM BTE solutions more clearly. Differences between the two approaches first increase as $Kn_x$ increases and then decrease as $Kn_x$ continues to increase towards the ballistic limit. The maximum error in the Fourier's Law solution occurs at $Kn_x \approx 0.5$ and is about 28%. Similar behavior is observed with and without internal heat generation, but the maximum error and the Knudsen number for which the maximum error occurs are seen to be problem specific.

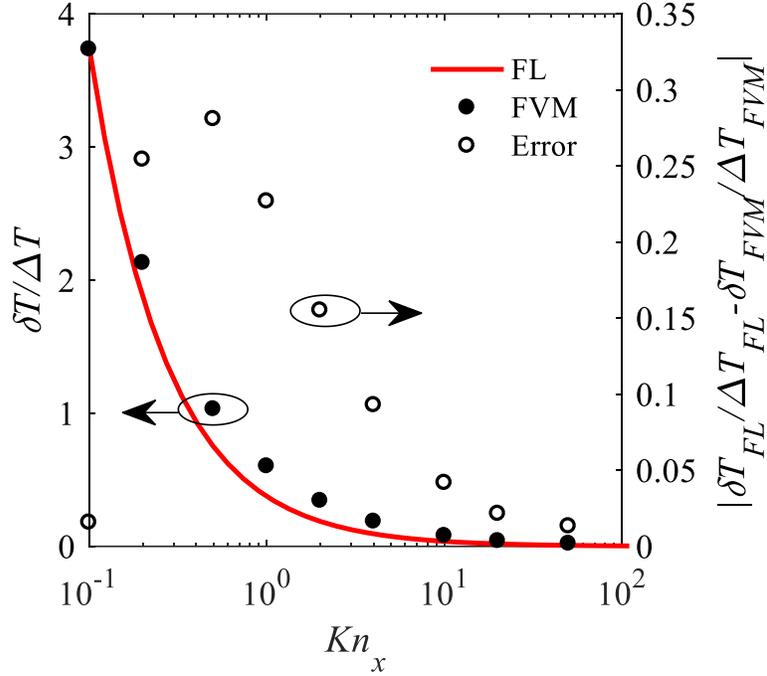

FIG. 7. The left axis shows the normalized temperature rise, $\delta T/\Delta T$ vs. $Kn_x$ for cross-plane thermal transport with internal heat generation (Fig. 1b). The line is the Fourier's Law solution from eqn. (20), and the filled symbols are FVM solutions of the phonon BTE. The empty symbols belong to the right axis and show the error, $|\delta T_{FL}/\Delta T_{FL} - \delta T_{FVM}/\Delta T_{FVM}|$, between both solutions.

**C. Apparent thermal conductivities**

Measuring internal temperature profiles is difficult experimentally; often what is determined is an apparent thermal conductivity. It is perhaps more relevant, therefore, to examine the errors associated with evaluating the apparent thermal conductivity with Fourier's Law. For case (a) in Fig. 1, a difference in the temperature between the two contacts with no internal heat generation, the apparent thermal conductivity that would be deduced was given by eqn. (15). Hua and Cao also define an apparent thermal conductivity for case (b) in Fig. 1, no temperature difference between the two contacts but with internal heat generation. In this case, the apparent thermal conductivity that would be deduced is [16]

$$\kappa_{app} = \frac{\dot{S} L_x^2}{12(\langle T(x) - T_0 \rangle)}, \tag{21a}$$

where

$$\langle T(x) \rangle = \frac{1}{L_x} \int_0^{L_x} T(x) dx. \tag{21b}$$

Using eqn. (18), we find

$$\kappa_{app} = \frac{\kappa_{bulk}}{1 + 4Kn_x}, \tag{22}$$

which is the same result obtained by Hua and Cao[16] with the ballistic-diffusive approach.[26] In the diffusive limit, $Kn_x \ll 1$, $\kappa_{app} \to \kappa_{bulk}$, as expected. As the structure becomes more ballistic, $\kappa_{app} < \kappa_{bulk}$, and in the ballistic limit where $Kn_x \gg 1$, $\kappa_{app} \to 0$.

Figure 8 plots the apparent thermal conductivities vs. Knudsen number for the case of no internal heat generation and for the case with internal heat generation. The Fourier's Law solutions, eqns. (15) and (22), are compared to FVM solutions to the phonon BTE. Again, we see that Fourier's Law is essentially exact in the diffusive and ballistic limits, and there is some error between these limits. For the apparent thermal conductivities, however, the errors are less than for the internal temperature profiles. The maximum error, $\Delta\kappa_{app}/\kappa_{bulk}$, is 5.6% for the results shown in Figs. 8 and 9. A properly implemented Fourier's Law, therefore, provides a good framework for interpreting measurements of apparent thermal conductivity.

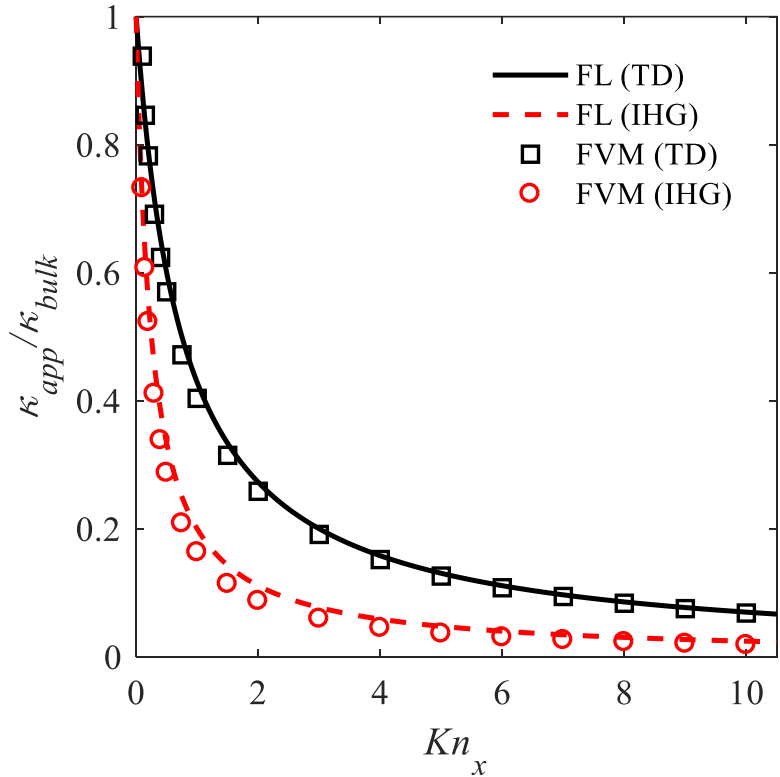

FIG. 8. Apparent thermal conductivities for cross plane thermal transport vs. $Kn_x$. Case of Fig. 1a (temperature difference but no internal heat generation (TD)) and case of Fig. 1b (no contact temperature difference but with internal heat generation (IHG)) are shown. Symbols are FVM simulations of the phonon BTE, and the solid lines are the Fourier's Law solutions, eqns. (15) and (22).

**D. Thin films**

We turn next to the thin films with diffuse boundary scattering. A proper treatment of these structures requires a two-dimensional solution. Extension of the methods described here to two and three dimensions is needed, but beyond the scope of this paper. Instead, we will examine one-dimensional (1D) solutions to these problems and show that 1D solutions can be quite accurate for the examples considered by Hua and Cao[16], who solved the 2D phonon BTE.

Following Hua and Cao, we examine the apparent thermal conductivity for the structures shown in Figs. 1c and 1d (additional comparisons to the Monte Carlo simulations of Hua and Cao are included in the Supplementary Information). Equation (15) gave the apparent thermal conductivity for the case of a temperature difference between contacts with no internal heat generation. In terms of the mean-free-path for backscattering in the bulk, $\lambda$, eqn. (15) can be written as

$$\kappa_{app} = \frac{C_V v_x^+ \lambda/2}{1 + \lambda/L_x}. \tag{23}$$

In a thin film, the mean-free-path is shortened by boundary scattering to

$$\frac{1}{\lambda_{TF}} = \frac{1}{\lambda} + \frac{1}{\beta L_y}, \tag{24}$$

where $\beta$ is an empirical parameter and $L_y$, the thickness of the film. Equation (24) can be regarded as an empirical fit to more rigorous treatments like that of Sondheimer[27] and McGaughey *et al.*[28] (See Supplementary Information for more discussion of this point.) Using (24) in (23) and expressing the result in terms of the Knudsen numbers $Kn_x = \Lambda/L_x$ and $Kn_y = \Lambda/L_y$, we find for the case of a temperature difference (TD),

$$\kappa_{app}(TD) = \frac{\kappa_{bulk}}{1 + \frac{4}{3}(Kn_x + Kn_y/\beta)}. \tag{25}$$

Equation (22) gave the apparent thermal conductivity for the case of no temperature difference between contacts with internal heat generation. In terms of the mean-free-path for backscattering in the bulk, $\lambda$, eqn. (22) can be written as

$$K_{app} = \frac{C_V v_x^+ \lambda/2}{1 + 3\lambda/L_x} \tag{26}$$

Using eqn. (24) for the mean-free-path in a thin film in eqn. (26) and expressing the result in terms of the Knudsen numbers $Kn_x = \Lambda/L_x$ and $Kn_y = \Lambda/L_y$, we find for the case of internal heat generation (IHG),

$$\kappa_{app}(IHG) = \frac{\kappa_{bulk}}{1 + \frac{4}{3}(3Kn_x + Kn_y/\beta)}. \tag{27}$$

We consider cases (c) and (d) of Fig. 1, transport in a thin film for $0.01 < Kn_x < 100$. Figure 9 compares the Fourier's Law and FVM BTE solutions for $Kn_y = 1$ assuming diffusive boundary scattering. (The apparent thermal conductivities for the TD and IHG cases are given by eqns. (25) and (27) for the Fourier's Law solution.) The TD and IHG apparent thermal conductivities are predicted by Fourier's Law to be distinctly different. Agreement between the FVM BTE and Fourier's Law solutions is quite good. The value, $\beta = 2.9$ in eqns. (25) and (27), which produces the best fit, is between the $3\pi/2$ given by Flik[29] and the 8/3 given by Majumdar[7].

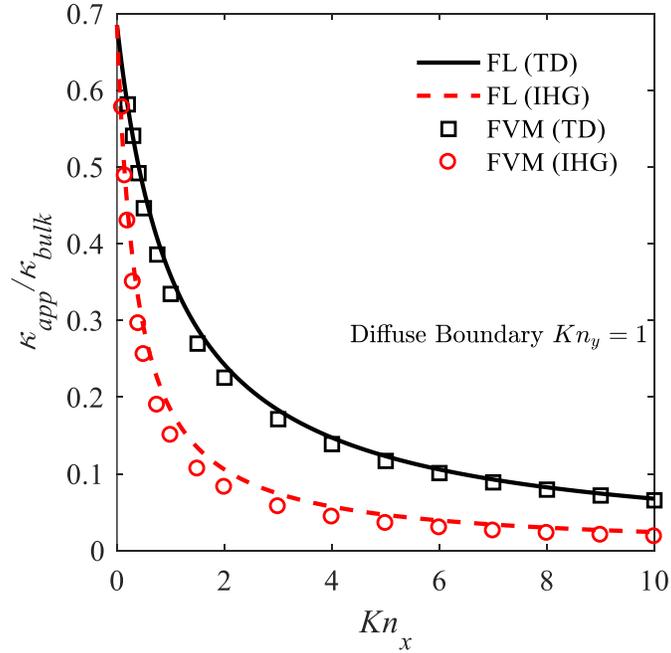

FIG. 9. Apparent thermal conductivities for a thin film with $Kn_y = 1$ vs. $Kn_x$ (cases (c) and (d) of Fig. 1). Symbols are FVM BTE simulation results, and the lines are the Fourier's Law solutions, eqns. (25) and (27), with $\beta = 2.9$.

**IV. DISCUSSION**

Several aspects of the solutions presented in the previous section are discussed in this section. First, we examine the directed temperatures, which play an important role in heat transport at the nanoscale.[12] Second, we examine the ballistic limit and show that the Fourier's Law solution has the correct ballistic limit. Third, we discuss the discrepancies observed between the Fourier Law and Monte Carlo solutions in the quasi-ballistic regime. Finally, we briefly discuss a recently reported, highly accurate analytical treatment of the problem with no internal heat generation.

**A. Directed temperatures and fluxes**

Figures 10 and 11 show the directed temperatures and heat fluxes for the cases of Figs. 1a and 1b – cross plane heat transport with and without internal heat generation. The directed temperatures are obtained from eqns. (12), and the corresponding directed fluxes from eqns. (3). As shown in Fig. 10a for the case with no internal heat generation, the forward flux is injected with the temperature of the left contact, $T_L$, and decays linearly across the film as phonon out-scattering takes place. Inside the film, the temperature, $T^+(x)$, should be regarded as a measure of the amount of heat in the forward flux. Similarly the reverse flux is injected at a temperature, $T_R$, and increases linearly across the film. The corresponding directed fluxes for this case are shown in Fig. 11a and follow directly from eqns. (1).

The case for internal heat generation is shown in Figs. 10b and 11b. As shown in Fig. 10b, $T^+(x)$ begins at $T_0$ and increases quadratically across the film as heat is generated. Similarly, $T^-(x)$ begins at $T_0$ at $x = L_x$ and increases across the film towards $x = 0$. The corresponding directed fluxes are shown in Fig. 11b. At $x = 0$, $F_Q^+(x = 0)$ begins at $F_0 = v_x^+ C_V T_0/2$, the heat flux injected from the contact. Similarly, at $x = L_x$, $F_Q^-(x = L_x)$ begins at $F_0$.

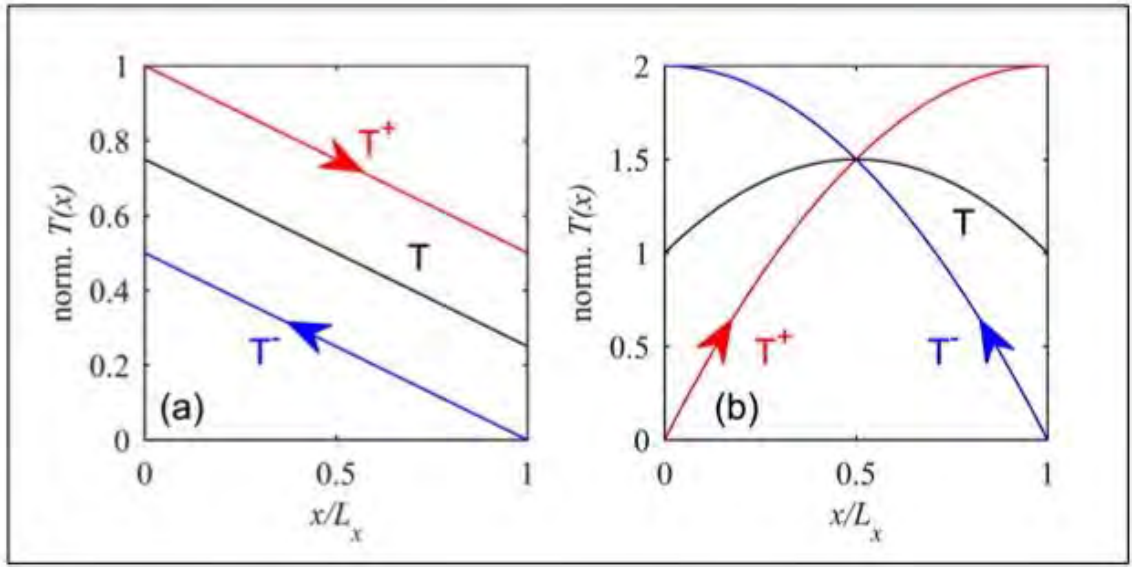

FIG. 10. Directed temperatures versus position $x/L_x$ for: a) Nanofilm (cross-plane) with temperature difference and b) Nanofilm (cross-plane) with internal heat source. In both cases, $L = \lambda = 4\Lambda/3 = 61.3$ nm. On the left, the normalized temperatures are defined as $T_{norm} = [T(x) - T_R]/[T_L - T_R]$. On the right, the normalized temperatures are $T_{norm} = [T(x) - T_0]/[\dot{S} L_x/2 v_x^+ C_v]$.

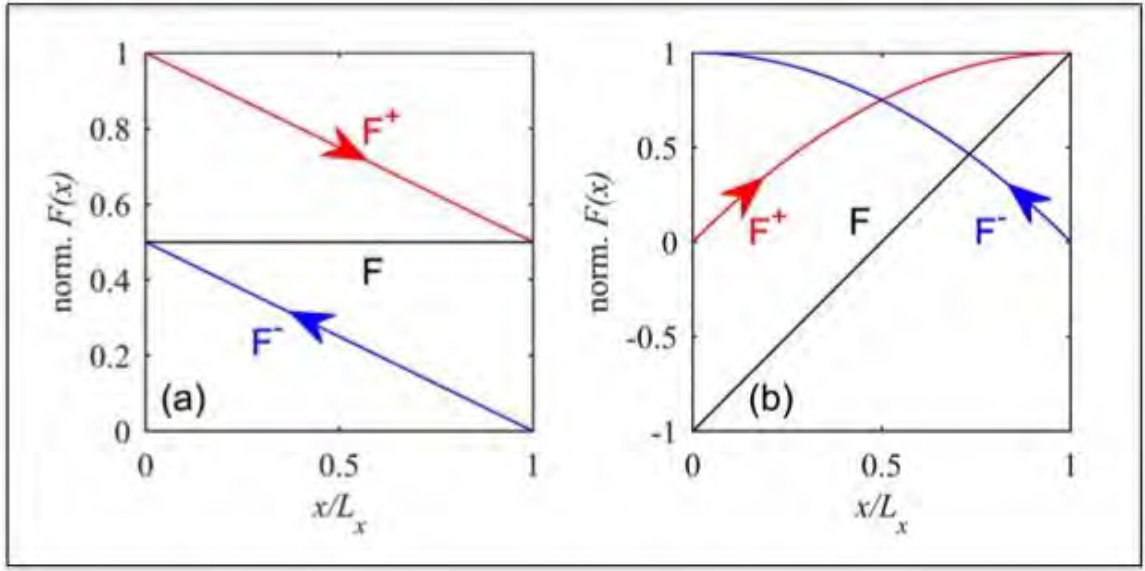

FIG. 11. Net flux and directed fluxes versus position $x/L_x$ for: a) Nanofilm (cross-plane) with temperature difference and b) Nanofilm (cross-plane) with internal heat source. In both cases, $L = \lambda = 4\Lambda/3 = 61.3$ nm. On the left, the normalized fluxes are defined as $F_{norm} = [F(x) - F^-(x = L_x)]/[F^+(x = 0) - F^-(x = L_x)]$. On the right, the normalized fluxes are $F_{norm} = [F(x) - F_0]/[\dot{S}L_x/2]$

## B. Ballistic limit

From the flux equations, (1), the ballistic limit is obtained by letting $\lambda \to \infty$. When converted to a temperature, the result is

$$T(x) = \frac{T_L + T_R}{2} + \frac{\dot{S}L_x}{2v_x^+ C_v}. \qquad (28)$$

For case (a) of Fig. 1, cross-plane thermal transport with no internal heat generation, we find $T(x) = (T_L + T_R)/2$, which is the correct ballistic limit.[12, 30] This is also the result obtained from the Fourier's law solution, eqns. (16) and (17) in the limit $\kappa_{bulk} \to \infty$. For case (b) of Fig. 1, cross-plane thermal transport with internal heat generation, $T_L = T_R = T_0$, and eqn. (23) gives the same result as the Fourier's Law solution, eqn. (18) in the limit as $\kappa_{bulk} \to \infty$. We conclude that Fourier's Law gives the correct solution in the ballistic and diffusive limits, but in between these limits, Figs. 2 and 4 show small differences between Fourier's Law and FVM solutions to the BTE.

## C. The quasi-ballistic regime, $Kn_x \sim 1$.

Fourier's Law gives correct solutions in the diffusive limit, and we have shown that when proper boundary conditions are used, it also gives the correct solutions in the ballistic limit, but as shown in Figs. 4 and 6, differences are observed in the quasi-ballistic regime where $Kn_x$ is on the order of unity. Under quasi-ballistic conditions, the temperature profiles in Fig. 3 are seen to be slightly non-linear – the temperature is a little higher than the Fourier Law results near the left contact and a little lower near the right contact. This nonlinearity can also be seen in Fig. 1 of Ref. 12 and in the exact solutions presented by Heaslet and Warming.[31] How is this explained?

A basic assumption in the flux method is that the forward flux and backward flux each travel at a fixed, spatially uniform velocity of $\langle v_x^+ \rangle = \langle v_x^- \rangle = v_s/2$. The factor of one-half comes from averaging over angles assuming a spherically symmetric distribution of velocities. It has, however, been noted that diffusion is altered within about a mean-free-path of absorbing contacts where the distribution function becomes asymmetric.[32] Berz has discussed this at the right (collecting) contact and Shockley at the left (injecting) contact.[19,33] This effect can be understood as follows. The heat flux is spatially invariant under the steady-state, no internal source conditions of Fig. 3. Write the heat flux as $F_Q = C_V T(x) \langle v_x(x) \rangle$, where $\langle v_x(x) \rangle$ is the average, $x$-directed phonon velocity at location, $x$. Near the right contact, the number of negative velocity phonons decreases, because the absorbing contact prevents their injection. As a result, the average velocity is larger than expected near the right contact, which requires the average temperature to be smaller than expected near the right contact to maintain the constant heat flux.[33] Near the left contact, the average velocity is smaller than expected because phonons with small $x$-directed velocities (i.e. those injected tangentially) scatter more often near the surface than do phonons with larger $x$-directed velocities.[19] Because the velocity is smaller than expected, the temperature must be larger than expected to maintain the constant heat flux. The distortion of the spherical distribution of velocities occurs within about a mean-free-path of each boundary. For very thin samples, these two regions overlap, and the error in our Fourier Law solution, which assumes a spherical distribution of velocities, is largest, as observed in Fig. 4. Similar distortions of the spherical distribution must explain the errors in the case of internal heat generation (Figs. 6 and 7). The boundary layer effects are resolved in full numerical[6] or analytical[10,31] solutions to the phonon BTE.

Finally, we note that when the contacts are at different temperatures, the magnitude of the temperature jumps depends on the phonon transmission (Knudsen number). When the temperatures of the two contacts are identical, but there is internal heat generation, temperature jumps can also occur, but they do not depend on the phonon transmission. It has been pointed out that in the general case, internal heat generation and contacts at different temperatures, it is possible to eliminate the temperature jumps or to produce asymmetric temperature jumps.[34]

**D. Analytical Solutions of Ordonez-Miranda et al.**

Highly accurate analytical solutions for case (a) in Fig. 1 have recently been reported by Ordonez-Miranda et al.[35] Their approach resolves the boundary layer non-linearities mentioned above, and are very close to the FVM numerical solutions (the difference is less than 2%). Analytical solutions such as these are very useful, but they tend to be available only for a few specific problems. For other problems, Fourier's Law can be used with modest errors. For example, cases (b), (c), (d), and (e) in Fig. 1 are easily handled by Fourier's Law. An arbitrary heat generation source, S(x), can also be treated, and extensions to full phonon dispersion and energy dependent scattering are possible (as discussed and demonstrated in Ref. 12). While Fourier's Law is not a panacea (for example, it's not clear how to extend it to strongly 2D problems), it can play a useful role in analyzing thermal transport at the nanoscale. In Fig. 5 of Ref. 35, the authors present analytical solutions for three different geometries. We discuss the corresponding Fourier Law solutions in the Supplementary Information.

## V. CONCLUSIONS

The results discussed in this paper show that when used with proper boundary conditions, the unmodified Fourier's Law can provide a good description of steady state, one-dimensional heat transport in nanostructures with and without internal heat generation (within the context of the simple gray model employed here). The results agree well (although not perfectly) with numerical solutions of the phonon BTE. They also agree very well with a more analytically complicated ballistic-diffusive approach.[26] The Fourier's Law approach provides simple, analytical expressions that are exact in the diffusive and ballistic limits. Between these two limits, errors in the Fourier's Law solution can occur. The problems discussed in this paper (and the additional ones in the Supplementary Information) indicate the magnitude of the errors that can be expected. For the apparent thermal conductivity, which can be measured more easily than the internal temperature profile, the errors are well below 10%.

The results of this paper also provide some insights into thermal transport at the nanoscale. For example, it is interesting to note that the magnitude of the temperature jump is related to the mean-free-path when there is no heat source, but it is independent of mean-free-path when there is an internal heat source and the contacts are at the same temperatures. We also showed how to extract the directed temperatures, $T^+(x)$ and $T^-(x)$ from $T(x)$. The results shown in Figs. 10 and 11 give insights into the meaning of temperature at the nanoscale; they show how it can be understood in a manner that is analogous to the way that electrochemical potentials at the nanoscale are now understood.[15]

To solve a heat transport problem, a heat current equation (e.g. Fourier's Law) is inserted into a heat balance equation, and boundary conditions are specified. This paper reinforces the conclusions of Refs. 10 and 12 that the main issue is not the validity of Fourier's Law at the nanoscale; it is the appropriate boundary conditions on the heat equation at the nanoscale.

Several issues deserve further study. A formal derivation of the flux equations from the phonon BTE would help to clarify the assumptions involved (a simple derivation is presented in the Supplementary Information). The Fourier's Law treatment of complex phonon dispersions and energy-dependent mean-free-paths deserves further study to extend the initial demonstration in Ref. 12. Extensions of this method to higher spatial dimensions should also be explored, but there are concerns about the usefulness of the diffusion approximation with temperature jumps in two and three-dimensions (see the discussion in Chapter 7, Sec. 6 of Ref. 3). Nevertheless, the in-plane transport examples discussed in the paper show that there are 2D problems for which a 1D approach is useful. We conclude that the results presented here support earlier suggestions that Fourier's Law can play a useful role in analyzing heat transport at the nanoscale.[12-14] More generally, this paper indicates how electron and phonon transport at the nanoscale can be understood within a common conceptual framework.[36]

**SUPPLEMENTARY MATERIAL**

See supplementary material for further explanations.

**ACKNOWLEDGEMENT**

This work was supported in part through the NCN-NEEDS program, which is funded by the National Science Foundation, contract 1227020-EEC, and by the Semiconductor Research Corporation and partially by the DARPA MATRIX program. JM also acknowledges partial support from NSERC of Canada. The authors benefitted from insightful discussions with T. S. Fisher of Purdue University and thank J. Y. Murthy of UCLA and D. Singh for providing the FVM solver for the phonon BTE.

**REFERENCES**

[1]D. G. Cahill *et al.,* "Nanoscale thermal transport II: 2003-2012," *Applied Physics Reviews*, **1**, p. 11305, 2014.

[2]D. G. Cahill *et al.,* "Nanoscale thermal transport," *J. Appl. Phys.*, **93**, pp. 793–818, 2003.

[3]G. Chen, N*anoscale Energy Transport and Conversion*, Oxford Univ. Press, Oxford, UK, 2005.

[4]A. A. Joshi and A. Majumdar, "Transient ballistic and diffusive phonon heat transport in thin films," *J. Appl. Phys.*, **74**, pp. 31–39, 1993.

[5]S.G. Volz and G. Chen, "Molecular dynamics simulation of thermal conductivity of silicon crystals," P*hys. Rev. B*, **61**, pp. 2651-2656, 2000.

[6]J.Y. Murthy and S.R. Mathur, "Computation of sub-micron thermal transport using an unstructured finite volume grid," Proc. Intern. Mech. Engineering Congress and Exposition (IMEC2001), pp. 1-8, 2001.

[7]A. Majumdar, "Microscale heat conduction in dielectric thin films," *J. Heat Transfer*, **115**, pp. 7-16, 1993.

[8]G. Chen and T. Zeng, "Nonequilibrium phonon and electron transport in heterostructures and superlattices," M*icroscale Thermophysical Engineering*, **5**, pp. 71-88, 2001.

[9]A.T. Ramu and Y. Ma, "An enhanced Fourier law derivable from the Boltzmann transport equation and a sample application in determining the mean-free=path of nondiffusive phonon modes," *J. Appl. Phys*., **116**, 093501, 2014.

[10]J.-P. M. Peraud and N. G. Hadjiconstantinou, "Extending the range of validity of Fourier's law into the kinetic transport regime via asymptotic solution of the phonon Boltzmann transport equation", *Phys. Rev. B*, **93**, 045424 (2016).

[11]C. Chen, Z. Du, and L. Pan, "Extending the diffusion approximation to the boundary using an integrated diffusion model, *API Advances,* **5**, 067115, 2015.

[12]J. Maassen and M. Lundstrom, "Steady-state heat transport: Ballistic-to-diffusive with Fourier's law," *J. Appl. Phys.*, **117**, p. 35104, 2015.

[13]J. Maassen and M. Lundstrom, "A simple Boltzmann transport equation for ballistic to diffusive transient heat transport," *J. Appl. Phys.*, **117**, p. 135102, 2015.

[14]J. Maassen and M. Lundstrom, "Modeling ballistic effects in frequency-dependent transient thermal transport using diffusion equations," *Journal of Applied Physics*, **119**, p. 95102, 2016.

[15]M. J. McLennan, Y. Lee, and S. Datta, "Voltage drop in mesoscopic systems: A numerical study using a quantum kinetic equation*," Phys. Rev. B*, **43**, pp. 13846 - 13884, 1991.

[16]Yu-Chao Hua and Bing-Yang Cao, "The effective thermal conductivity of ballistic–diffusive heat conduction in nanostructures with internal heat source," *International Journal of Heat and Mass Transfer*, **92,** pp. 995–1003, 2016.

[17]D. Singh and J.Y. Murthy, "Phonon BTE Finite Volume Solver," private communication, July 29, 2016.

[18]Yu-Chao Hua and Bing-Yang Cao, "Phonon ballistic-diffusive heat conduction in silicon nanofilms by Monte Carlo simulations," *International Journal of Heat and Mass Transfer*, **78**, pp. 755–759, 2014

[19]W. Shockley, "Diffusion and drift of minority carriers in semiconductors for comparable capture and scattering mean free paths," *Phys. Rev,* **125,** no. 5, 1570–1576, 1962.

[20]J. P. McKelvey, R. L. Longini, and T. P. Brody, "Alternative approach to the solution of added carrier transport problems in semiconductors," *Phys. Rev,* **123**, pp. 51–57, 1961.

[21]Changwook Jeong, Raseong Kim, Mathieu Luisier, Supriyo Datta, and Mark Lundstrom, "On Landauer vs. Boltzmann and full band vs. effective mass evaluation of thermoelectric transport coefficients," *J. Appl. Ph*ys., **107**, 023707, 2010.

[22]D.B. Olfe, "A modification of the differential approximation for radiative transfer," *AIAA Journal*, **5**, pp. 638-643, 1967.

[23]A.M. Gheitaghy and M.R. Talaee, "Solving hyperbolic heat conduction using electrical simulation," *J. Mechanical Sci. and Technology*, **27**, pp. 3885-3891, 2013.

[24]T. Zeng and G. Chen, "Phonon heat conduction in thin films: Impacts of thermal boundary resistance and internal heat generation," *Trans. ASME*, **123**, pp. 340-347, 2001.

[25]A. Balusu and D.G. Walker, "One-dimensional thin-film phonon transport with generation," *Microelectronics Journal*, **39**, pp. 950-956, 2008.

[26]G. Chen, "Ballistic-diffusive heat-conduction equations," *Phys. Rev. Lett.,* **86**, pp. 2297–2300, 2001.

[27]E. H. Sondheimer, "The mean free path of electrons in metals," *Advances in Physics*, **50**, pp. 499–537, 1952.

[28]A.J.H. McGaughey, E.S. Landry, D.P. Sellan, and C.H. Amon, "Size-dependent model for thin-film and nanowire thermal conductivity," *Appl. Phys. Lett*., **99**, 131904, 2011.

[29]M.L. Flik and C.L. Tien, "Size effect on the thermal conductivity of high-$T_c$ thin-film superconductors," *J. Heat Transfer*, **112**, pp. 872-881, 1990.

[30]Gang Chen, "Particularities of heat conduction in nanostructures," *Journal of Nanoparticle Research,* **2**, pp. 199–204, 2000.

[31]M.A. Heaslet and R.F. Warming, "Radiative transport and wall temperature slip in an absorbing planar medium," *Int. J. Heat Mass Transfer*, **8**, pp. 979-994, 1965.

[32]G. Baccarani, C. Jacoboni, and A.M. Mazzone, "Current transport in narrow-base transistors," *Solid-State Electronics*, **20**, pp. 5-10, 1977.

[33] F. Berz, "Diffusion near an absorbing boundary," *Solid-State Electronics*, **17**, pp. 1245-1255, 1974.

[34] A.A. Candadai and Vaibhav Jain (private communication, June, 2016).

[35] J. Ordonez-Miranda, R. Yang, S. Volz,, and J.J. Alvarado-Gil, "Steady-state and modulated heat conduction in layered systems predicted by the analytical solution of the phonon Boltzmann transport equation," *J. Appl. Phys.*, **118**, 075103, 2015.

[36] S. Datta, *Lessons from Nanoelectronics: A New Perspective on Transport*, Vol. 1 in Lessons from Nanoscience: A Lecture Notes Series, World Scientific, 2012.